\documentclass[prb,twocolumn]{revtex4}
\usepackage[dvips]{graphicx}
\usepackage{latexsym,amsmath,amssymb,bm,euscript,eufrak}

\def\etal{~\textit{et~al.}} 
\def\ra{\rangle} 
\def\la{\langle} 
\def\up{\uparrow}
\def\dn{\downarrow}
\def\Hc{{\rm H.c.}}
\def\cc{{\rm c.c.}}
\def\ET{{$\kappa$-(ET)$_2$Cu$_2$(CN)$_3$}}
\def\bondpic{{\begin{picture}(17,6)(-2,-2)
	            \put (0,0) {\circle*{5}}
	            \put (12,0) {\circle*{5}}
		    \put (0,0) {\line (1,0) {12}}
              \end{picture}}}
\def\rhombpic{{\begin{picture}(26,15)(-2,-2)
                     \put (0,0) {\circle*{5}}
		     \put (12,0) {\circle*{5}}
		     \put (6,10) {\circle*{5}}
		     \put (18,10) {\circle*{5}}
		     \put (0,0) {\line (1,0) {12}}
		     \put (6,10) {\line (1,0) {12}}
		     \put (0,0) {\line (3,5) {6}}
		     \put (12,0) {\line (3,5) {6}}
               \end{picture}}}
\def\triangpic{{\begin{picture}(17,15)(-2,-2)
                      \put (0,0) {\circle*{5}}
		      \put (12,0) {\circle*{5}}
		      \put (6,10) {\circle*{5}}
		      \put (0,0) {\line (1,0) {12}}
		      \put (12,0) {\line (-3,5) {6}}
		      \put (0,0) {\line (3,5) {6}}
                \end{picture}}}
\def\looptripic{{\begin{picture}(17,15)(-2,-2)
                       \put (0,0) {\circle*{5}}
		       \put (12,0) {\circle*{5}}
		       \put (6,10) {\circle*{5}}
		       \put (6,3) {\circle{14}}
                 \end{picture}}}
\def\loopfourpic{{\begin{picture}(15,15)(-2,-2)
                      \put (0,0) {\circle*{5}}
		      \put (10,0) {\circle*{5}}
		      \put (10,10) {\circle*{5}}
                      \put (0,10) {\circle*{5}}
		      \put (5,5) {\circle{14}}
                   \end{picture}}}
\def\JJJ{{$J_2$-$J_3$-$J_4$}}
\def\Phiext{\Phi^{\rm ext}}
\def\phiint{\phi^{\rm int}}
\def\tspinon{t^{\rm spinon}}

\begin{document}

\title{Orbital magnetic field effects in spin liquid with spinon 
Fermi sea: \\
Possible application to $\kappa$-(ET)$_2$Cu$_2$(CN)$_3$}
\author{Olexei I. Motrunich}
\affiliation{Kavli Institute for Theoretical Physics, 
University of California, Santa Barbara, CA 93106-4030}

\date{October 23, 2005}

\begin{abstract}
We consider orbital magnetic field effects in a spin liquid phase of a 
half-filled triangular lattice Hubbard system close to the Mott
transition, continuing an earlier exploration of a state with 
spinon Fermi surface.
Starting from the Hubbard model and focusing on the insulator side, 
we derive an effective spin Hamiltonian up to four-spin exchanges 
in the presence of magnetic field, and find that the magnetic 
field couples linearly to spin chirality on the triangles.  
The latter corresponds to a flux of an internal gauge field in a
gauge theory description of the spin liquid, and therefore a static 
such internal flux is induced.
A quantitative estimate of the effect is obtained using a spinon mean 
field analysis, where we find that this orbital field seen by the spinons
is comparable to or even larger than the applied field.
We further argue that because the stiffness of the emergent internal 
gauge field is very small, such a spinon-gauge system is strongly 
susceptible at low temperatures to an instability of the 
homogeneous state due to strong Landau level quantization for spinons.
This instability is reminiscent of the so-called strong magnetic
interaction regime in metals with the usual electromagnetic field,
but we estimate that the corresponding temperature--magnetic field 
range is significantly broader in the spinon-gauge system.
\end{abstract}

\maketitle

\section{Introduction}
Experimental studies\cite{ET_SL1, ET_SL2, Kanoda_kitp} of the 
quasi-two-dimensional organic material \ET\ strongly suggest a 
spin liquid state in the insulating phase at ambient pressure.
\ET\ is a strongly correlated half-filled Hubbard system on an almost 
isotropic triangular lattice.  The material is just near the boundary of 
the insulator-metal transition.\cite{ET_SL1, ET_SL2, Kanoda_kitp, Komatsu, Kawamoto, McKenzie, Imada}  
It is an insulator at ambient pressure with a charge gap of order
$350$~K, and shows no signs of magnetic ordering down to $32$~mK 
despite a relatively large exchange coupling $J \sim 250$~K.
Experiments find that this spin liquid maintains a finite 
susceptibility and a finite $1/(T_1 T)$ for the $^{13}$C nuclear spin 
relaxation rate down to low temperatures,\cite{ET_SL1, Kawamoto} 
and also that this insulator has large spin entropy.\cite{ET_SL2, Kanoda_kitp}  

These observations and related numerical studies\cite{Imada, LiMing}
led the present author,\cite{ringxch1} 
and also the authors of Ref.~\onlinecite{SSLee}, to propose a spin liquid
state with spinon Fermi surface as a likely candidate.
In Ref.~\onlinecite{ringxch1}, this state was shown to be
energetically favored in a spin model with Heisenberg and ring
exchanges appropriate for the description of the insulator,
while Ref.~\onlinecite{SSLee} used the mean field slave particle
approach and also derived an effective gauge theory description
of the proposed spin liquid.

Here, we examine the response of this spin liquid to strong
magnetic fields.
One motivation is to look for possible direct probes of the spinon
Fermi surface similar to the ones used in metals such as
magneto-oscillations or magnetoacoustic resonance.
Another motivation is the observations in 
Refs.~\onlinecite{Kanoda_kitp}~and~\onlinecite{Kawamoto} of
significant broadening of the $^{13}$C NMR line in a large magnetic 
field up to $9$~T at low temperatures below $10-30$~K.
Systematic studies\cite{Kanoda_kitp} reveal that the lines broaden 
symmetrically and the broadening increases with increasing field and 
also with decreasing temperature.  
At the present, it is not clear whether this anomalous behavior
is intrinsic, or is the result of extrinsic disorder effects in
the critical spin liquid.

A simple intuition about the Mott insulator would be that the charge 
motion is suppressed and only spin degrees of freedom couple to the 
magnetic field.  If this were the case, then one would not expect any 
strong intrinsic effects in the presence of the magnetic field.  
In particular, in the spin liquid state with spinon Fermi surface 
and just the Zeeman spin coupling to the applied field one would expect 
only Pauli spin paramagnetism.  
We argue, however, that orbital effects need to be carefully included 
when analyzing the response of the spin liquid in \ET.
This is because charge fluctuations, which become more prominent in the 
vicinity of the insulator-metal transition, also induce an effective 
``orbital'' coupling to the magnetic field.  
The effect of such coupling on the proposed spinon Fermi sea state is 
further amplified by the fact that this phase itself is stabilized
against other competing spin liquids or the antiferromagnetically 
ordered state by charge fluctuations that produce the four-spin 
ring exchange terms.\cite{ringxch1, SSLee}

First, we show that in the presence of the external magnetic 
field the spinons effectively experience an ``internal'' orbital field 
that is comparable to and maybe even larger than the applied field.  
This is despite the fact that spinons do not transport
electrical charge, but follows when we derive an effective spin 
Hamiltonian from electronic degrees of freedom in the presence of the
applied magnetic field.  We find that the magnetic field couples 
linearly to the spin chirality on the elementary triangles.
The effective description of the spin liquid state contains spinons 
coupled to a dynamically generated ``internal'' gauge field.
The physical meaning of the flux of the internal gauge field
is precisely the spin chirality, and the external magnetic field 
therefore induces a static internal flux seen by the spinons 
that is comparable to the applied field.

Second, we argue that because the stiffness of the internal gauge field 
is very small (see Appendix~\ref{app:gt} for numerical estimates),
the response of the spinon-gauge field system changes dramatically
at low temperatures such that the Landau quantization of the
spinons in the static internal field is not smeared by the temperature.
These effects are similar to strong electronic magnetism in quantizing 
field at low temperatures familiar in magneto-oscillation studies of 
metals\cite{Shoenberg, Abrikosov, Holstein}
(which we also review in Appendix~\ref{app:mint}).
In particular, the homogeneous state with continuously varying
internal field becomes unstable below some temperature, which for the 
\ET\ spin liquid we estimate to be several Kelvin in typical 
laboratory fields.
The instability regime is significantly wider in the spinon-gauge 
system than in metals because the internal gauge field stiffness is 
so much smaller than that of the physical electromagnetic field.  
Crudely, the spinon states with integer Landau level filling are
more stable than the states with a continuously varying filling, 
and it becomes advantageous for the internal gauge field to adjust 
itself discontinuously to achieve this.
This instability also preempts the possibility of a direct observation 
of the spinon Fermi surface using magneto-oscillation probes,
but can be viewed as an extreme manifestation of such 
magneto-oscillations.

The objective of the present paper is to characterize the above two
predictions in the spinon-gauge system focusing on the \ET\ material.  
The paper is organized as follows.
In Sec.~\ref{sec:Hring} (and also Appendix~\ref{app:Hring}),
we describe the effective spin Hamiltonian derived from the triangular 
lattice Hubbard model in the magnetic field.
In the main part of the paper, Sec.~\ref{sec:mf}, we perform
a spinon mean field study of the effective Hamiltonian.
We start with a review of the system in zero field\cite{ringxch1} 
and progressively add the needed ingredients for the discussion of the 
response to the magnetic field such as the appearance of the static 
internal gauge field and the importance of the spinon Landau level 
quantization at low temperatures.
We then argue that a homogeneous spin liquid state with continuously 
varying static internal flux becomes unstable at low but experimentally 
relevant temperatures of few Kelvin.  We finalize this section
with a more careful discussion of the physical setting in the real 
system.
Helpful connections with the gauge theory description and 
analogies with magnetic interaction effects in metals are summarized 
in Appendixes~\ref{app:gt}~and~\ref{app:mint}.
Throughout, we use the \ET\ parameters as a guide for the relevant 
questions.  In Sec.~\ref{sec:ET}, we collect the main estimates
for the \ET\ material and consider some experimental aspects.

\section{Ring exchange Hamiltonian in the presence of magnetic field}
\label{sec:Hring}
The approach adopted in Ref.~\onlinecite{ringxch1} and pursued here is 
to focus on the spin degrees of freedom when describing the insulating 
state.  This is achieved by considering an effective spin Hamiltonian 
obtained from the microscopic Hubbard model by a canonical transformation
that projects out the double occupancy.
The importance of the charge fluctuations is retained in the 
form of more complicated multispin exchanges.
The spin system is still more amenable to analysis because there
is much less disproportion between the relevant energy scales 
and the couplings in the effective Hamiltonian, unlike in the original 
Hubbard model.

The effective Hamiltonian to order $t^4/U^3$ on the isotropic triangular 
lattice reads (see also Appendix~\ref{app:Hring})
\begin{eqnarray}
{\hat H}_{\rm eff} &=& J_2 \sum_\bondpic P_{12}
+ J_4 \sum_\rhombpic (P_{1234} + \Hc) \\
&+& J^{\prime\prime} \sum_{\la\!\la ij \ra\!\ra} 
    {\bm S}_i \cdot {\bm S}_j
+ J^{\prime\prime\prime} \sum_{\la\!\la\!\la ij \ra\!\ra\!\ra}  
    {\bm S}_i \cdot {\bm S}_j \\
&+& \Phi_\triangle^{\rm ext} \,\, J_3 \sum_\triangpic 
    \,\,i\,\, (P_{123} - \Hc) ~.
\label{Heff}
\end{eqnarray}
Here we use multispin exchange operators defined as
$P_{12\dots n} : |\sigma_1, \sigma_2, \dots, \sigma_n \ra 
             \to |\sigma_n, \sigma_1, \dots, \sigma_{n-1} \ra$.
For two spins, this reduces to the familiar Heisenberg exchange,
$P_{12} = P_{12}^\dagger = 2 {\bm S}_1 \cdot {\bm S}_2 + \frac{1}{2}$.
The first two lines give the effective Hamiltonian in the absence of the
magnetic field.  As discussed in Ref.~\onlinecite{ringxch1}, 
important terms in this Hamiltonian are the nearest-neighbor 
two-spin exchanges and the ring exchanges around the rhombi of the 
triangular lattice.  The corresponding coupling constants are given in 
terms of the Hubbard model parameters as
\begin{equation}
J_2 = \frac{2t^2}{U} \left(1-\frac{32t^2}{U^2}\right) ~, \quad\quad
J_4 = \frac{20t^4}{U^3} ~.
\label{J2J4}
\end{equation}
The second- and third-neighbor Heisenberg exchanges 
$J^{\prime\prime} = -16t^4/U^3$ and $J^{\prime\prime\prime} = 4t^4/U^3$, 
though nominally of the same order as the ring exchange coupling, 
were argued to play a minor role because of the weak correlations 
between such further-neighbor spins in the magnetically disordered
state.

The last line in Eq.~(\ref{Heff}) shows a new term that appears
at order $t^3/U^2$ in the presence of the magnetic field.  
This term involves three-spin exchanges around the elementary triangles 
and is proportional to the enclosed flux with the coupling constant
\begin{equation}
J_3 = \frac{6t^3}{U^2} ~.
\end{equation}
The dimensionless flux is 
$\Phiext_\triangle = e B A_\triangle/(\hbar c)$,
where $B$ is the field and $A_\triangle$ is the triangle area.
The Hamiltonian Eq.~(\ref{Heff}) is written to linear order in the 
external flux, which is assumed to be small, $\Phiext_\triangle \ll 1$.
Each triangle is counted once and is traversed in the same direction.
We can write the $J_3$ term more explicitly as
$\sin(\Phiext_{123}) \,i\, (P_{123}-\Hc)$, where $\Phiext_{123}$ is
the complex phase of the loop product $t_{12} t_{23} t_{31}$ of the 
electron hopping amplitudes in the field.
In general, each contribution from an exchange path that encloses 
flux is affected by the magnetic field (cf.~Appendix~\ref{app:Hring}).
However, the couplings $J_2$, $J_4$, $J^{\prime\prime}$, and
$J^{\prime\prime\prime}$, are modified only at quadratic order
in $\Phiext$.  The exhibited three-spin terms represent the full
effect linear in $\Phiext$ to order $t^4/U^3$.
The effective Hamiltonian contains only terms $P_{12}$,
$(P_{1234}+\Hc)$, and $\,i\,(P_{123}-\Hc)$, which was achieved
with the help of the following identities valid in the spin-1/2 case:
\begin{eqnarray}
\label{ReP3}
P_{123} + \Hc &\!=\!& P_{12} + P_{23} + P_{31} - 1 ~, \\
\label{ImP4}
P_{1234} - \Hc &\!=\!& \frac{1}{2}
\left(P_{123} + P_{234} + P_{341} + P_{412} - \Hc \right).
\end{eqnarray}

The three-spin operator $\,i\,(P_{123}-\Hc)$ has a simple physical 
meaning -- it represents the spin chirality:
\begin{equation}
i\, (P_{123} - \Hc) = 
-\,4\, {\bf S}_1 \cdot {\bf S}_2 \times {\bf S}_3 ~.
\end{equation}
Thus, the external field couples linearly to the spin chirality
and therefore induces such chirality density in the system.

In the above discussion, we have not mentioned the original
Zeeman coupling of the electrons to the magnetic field.
Since the Zeeman term involves the conserved total $S_z$,
we can simply reinstate it in the final effective spin
Hamiltonian.

For the \ET, we use $t=55~$meV and $U/t = 8.2$
to obtain $J_2 \simeq 7~$meV, $J_4/J_2 \simeq 0.3$,
and $J_3/J_2 \simeq 0.7$;
the Heisenberg exchange coupling is $J=2J_2 \simeq 160~$K,
which is comparable with the estimate $J \simeq 250~$K in 
Ref.~\onlinecite{ET_SL1}.

To summarize, when charge fluctuations are significant,
the effective spin Hamiltonian in the presence of the applied 
magnetic field is modified beyond the direct Zeeman term and 
contains a linear coupling of the spin chirality to the external 
field.  This orbital effect is important when considering the response 
of such insulator to the magnetic field.

\section{Spinon mean field description}
\label{sec:mf}

\subsection{General setting}
In Ref.~\onlinecite{ringxch1} we considered the situation with no
magnetic field and argued that the four-spin ring exchanges stabilize
the spin liquid state with spinon Fermi surface.
This state can be viewed as a Gutzwiller projection of a 
fermionic mean field state in which spinons hop on the triangular
lattice.  We performed a direct variational wave function
study, and also provided an intuitive mean field argument for the 
stabilization of this state.

A similar mean field treatment is pursued here to understand the effects 
of the magnetic field.  To simplify the discussion, we will focus
on the \JJJ\ terms in $H_{\rm eff}$.  As mentioned earlier, the second- 
and third-neighbor Heisenberg exchanges are not expected to have 
significant effect.

Each spin-1/2 is represented in terms of two ``spinon'' operators
$f_\up, f_\dn$, with the occupancy constraint\cite{LeeNagaosaWen, WenPSG}
\begin{equation}
f_{r \sigma}^\dagger f_{r \sigma} = 1 ~.
\end{equation}
In this representation, the \JJJ\ Hamiltonian reads
\begin{eqnarray*}
& \hat H & = 
J_2 \sum_{\rm links}
(f_{1\alpha}^\dagger f_{1\beta}) (f_{2\beta}^\dagger f_{2\alpha}) \\
\!\!&+&\!\! J_4 \!\!\! \sum_{\rm rhombi} \!\!
\left[
(f_{1\alpha}^\dagger f_{1\beta}) (f_{2\beta}^\dagger f_{2\gamma})
(f_{3\gamma}^\dagger f_{3\delta}) (f_{4\delta}^\dagger f_{4\alpha})
+ \Hc \right] \\
\!\!&+&\!\! J_3 \!\!\! \sum_{\rm triangles} \!\!\!
\sin(\Phiext_{123}) \,i\,
\left[
(f_{1\alpha}^\dagger f_{1\beta}) (f_{2\beta}^\dagger f_{2\gamma})
(f_{3\gamma}^\dagger f_{3\alpha}) 
- \Hc \right] ~.
\end{eqnarray*}
For clarity, we use the form of the $J_3$ term valid for general 
$\Phiext$ (cf.~Appendix~\ref{app:Hring}), even though we are interested 
in the case of small $\Phiext_\triangle \ll 1$.

We consider trial spinon hopping Hamiltonians\cite{WenPSG, DiD}
\begin{eqnarray}
\hat H_{\rm trial} \!=\!  
-\! \sum_{\la rr' \ra} \!\!\left( 
  \tspinon_{rr'} f_{r\alpha}^\dagger f_{r'\alpha} + \Hc 
                             \right)
\!-\! \sum_r \mu_r f_{r\alpha}^\dagger f_{r\alpha}~,
\label{Htrial}
\end{eqnarray}
with general hopping amplitudes $\tspinon_{rr'}=(\tspinon_{r'r})^*$.  
The occupancy constraints are implemented on average by the appropriate 
chemical potentials $\mu_r$.
The mean field scheme that we use here is to evaluate the above 
exchange terms by contracting only the fermions with the same spin index.
This approach can be justified in a large-$N$ fermionic generalization, 
since any other contraction is down by a factor of $1/N$.  
While the present $N=2$ is not large, the other contractions
do not introduce qualitatively new terms, and this scheme
is expected to capture the relevant physics rather well.
The mean field energy is
\begin{eqnarray}
\label{genEmf}
E_{\rm mf} & = & -4 g_2 J_2 \sum_{\rm links} |\chi_{12}|^2 \\
&-& 16 g_4 J_4 \sum_{\rm rhombi}
             \big[\chi_{12}\chi_{23}\chi_{34}\chi_{41} + \cc \big] 
\nonumber \\
&-& 8 g_3 J_3 \sum_{\rm triangles} \sin(\Phiext_{123}) 
             (-i) \big[\chi_{12}\chi_{23}\chi_{31} - \cc \big] ~.
\nonumber
\end{eqnarray}
Here $\chi_{rr'} \equiv \la f_{r'}^\dagger f_r \ra$ are link
expectation values evaluated for one spin species, and the 
powers-of-two numerical factors appear from the spin summations.
We also introduce phenomenological renormalization factors
$g_2$, $g_4$, and $g_3$, to keep us aware of the schematic character 
of the mean field treatment.
In a more quantitative treatment, these factors can be estimated 
by matching to numerical evaluations with the Gutzwiller-projected 
wave functions.
This is done in Appendix~\ref{app:grenorm}, where we find that
$g_2 \simeq 1.7$, $g_4 \simeq 8$, and $g_3 \simeq 5$, reproduce
fairly well such direct trial wave function computations.

The mean field energy is to be minimized over the trial spinon
hopping amplitudes $\tspinon_{rr'}$.
We focus on the states that describe translationally invariant 
spin liquids.\cite{WenPSG}  These are the projected Fermi sea state with
real hopping amplitudes and the so-called flux states with complex 
$\tspinon_{rr'}$ realizing nontrivial ``internal'' fluxes through the 
hopping loops.  
The flux can be either uniform or have a staggered pattern.
The projected Fermi sea state is a special case with zero flux.
Here, we are primarily interested in the uniform flux states since
these are natural candidates in the presence of the external magnetic 
field if one starts with the zero-flux state in the absence of the
magnetic field.

The mean field energy per site for a uniform flux state is
\begin{eqnarray}
\epsilon_{\rm mf} & = & -12 g_2 J_2 |\chi_\phi|^2
- 96 g_4 J_4 |\chi_\phi|^4 \cos(\phiint_{1234}) \nonumber \\
&-& 32 g_3 J_3 |\chi_\phi|^3 \sin(\Phiext_{123}) \sin(\phiint_{123}) ~.
\label{emfphi}
\end{eqnarray}
Here, $\chi_{rr'} = |\chi_\phi| \exp(i a_{rr'})$ and $\phiint$ is 
the flux of the ``internal'' gauge field $a_{rr'}$.  
Note that the link expectation values $\chi_{rr'}$ obtain the same 
flux pattern as the input amplitudes $\tspinon_{rr'}$.
The gauge theory language is explained in Appendix~\ref{app:gt}.

\subsection{Review of the zero-field case}
\begin{figure}
\centerline{\includegraphics[width=3.0in]{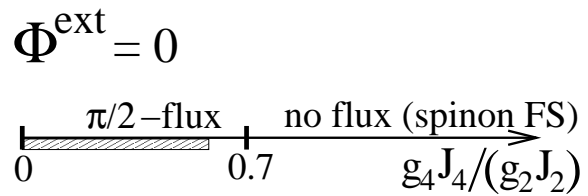}}
\vskip -2mm
\caption{
``Phase diagram'' from the mean field energy optimization over 
translationally invariant states in the absence of the magnetic field.  
In the hatched region, the zero-flux state is also unstable against 
introducing small internal flux.  
For the \ET\ parameters, we estimate 
$g_4 J_4/(g_2 J_2) \simeq 1.2 - 1.5$.
}
\label{fig:mfphased}
\end{figure}

Let us first consider $\Phiext=0$ following Ref.~\onlinecite{ringxch1}. 
We find that for small $g_4 J_4/(g_2 J_2) \leq 0.69$, 
the lowest energy state has $\phi_\triangle=\pi/2$ flux through each 
triangle, while for larger ring exchanges the best state has zero flux
(see Fig.~\ref{fig:mfphased}).
The corresponding numerical values of $|\chi_\phi|$ for the half-filled
triangular lattice are
\begin{eqnarray}
|\chi_{\phi=\pi/2}| = 0.2002, \quad\quad
|\chi_{\phi=0}| = 0.1647 ~.
\end{eqnarray}
It is known\cite{flux_states} that flux states have large absolute value 
$|\chi_\phi|$ and therefore good Heisenberg energies.  
On the other hand, the ring exchanges are directly sensitive to
the placket fluxes and dislike any fluxes, as can be seen from 
Eq.~(\ref{emfphi}).  This is why the zero-flux state is stabilized for 
large $J_4/J_2$.  As explained in Ref.~\onlinecite{ringxch1},
we do not consider so-called dimer states even though formally
for fixed $g_2 J_2$ these have the lowest mean field energy in the range
$g_4 J_4/(g_2 J_2) \leq 2.4$.  Our reasoning is that $J_2$ is fixed,
and to go from the mean field evaluations to the projected 
wave functions, the renormalization factor $g_2$ is larger for the 
translationally invariant states.

\begin{figure}
\centerline{\includegraphics[width=\columnwidth]{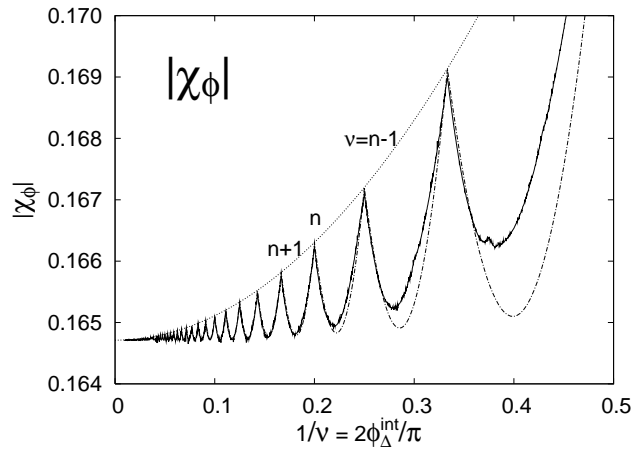}}
\vskip -2mm
\caption{
Mean field numerical data for uniform flux states on the half-filled 
triangular lattice.  
The figure shows the absolute value of the bond expectation value 
$|\chi_\phi| \equiv |\la f_{r'}^\dagger f_r \ra|$ as a
function of the spinon filling factor $\nu$ with respect to the
internal flux (one spin species is considered).
The dotted line is the enveloping function Eq.~(\ref{chiphi_envelope}),
while the dash-dotted line is the full model function
Eq.~(\ref{chiphi_model}).
The data plotted correspond to rather large $\phiint$, but the same 
behavior is expected to continue for small fluxes.
}
\label{fig:chiphi}
\end{figure}

For later convenience, Fig.~\ref{fig:chiphi} shows the behavior of
$|\chi_\phi|$ evaluated numerically for the triangular lattice flux 
states.  It is useful to note the enveloping function for small fluxes,
which we find to be 
\begin{eqnarray}
|\chi_\phi| \leq |\chi_0| (1 + c \phi_\triangle^2)
\label{chiphi_envelope}
\end{eqnarray}
with $c \approx 0.1$.
In particular, it follows that for small $g_4 J_4 / (g_2 J_2) \leq 0.56$
-- hatched region in Fig.~\ref{fig:mfphased} --
the zero-flux state is also unstable against introducing weak fluxes 
(while the $\pi/2$-flux state is the global translationally-invariant
minimum till a somewhat larger value of $0.69$).

A direct Gutzwiller wavefunction study gives that the 
zero-flux state has lower energy than the $\pi/2$-flux state for 
$J_4 \geq 0.145 J_2$.  From this and similar comparisons we estimate
that $g_4/g_2$ is roughly 4 to 5 -- see Appendix~\ref{app:grenorm}
for more details.

In the \ET\ compound, we have $J_4/J_2 \simeq 0.3$, which is not far
from the regime where it would be advantageous to spontaneously generate 
such internal flux.  
We therefore reason that the spinon Fermi sea 
state in the \ET\ compound is rather susceptible to the internal
flux generation.  In particular, we expect some enhancement in the
response to the external magnetic field, to which we now turn.

\subsection{Mean field over homogeneous flux states in the presence
of the magnetic field}
Consider the mean field energy Eq.~(\ref{emfphi}) with nonzero
but small $\Phiext$.  Let us first try the enveloping function 
Eq.~(\ref{chiphi_envelope}) for small $\phiint$.  
Expanding $\epsilon_{\rm mf}$ to quadratic
order in $\phiint$ and minimizing over the internal flux,
we find
\begin{equation}
\phiint = \frac{g_3 J_3}
{12 g_4 J_4 |\chi_0| 
 \Big[1 - 2c\left(1 + \frac{g_2 J_2}{16 g_4 J_4 |\chi_0|^2}\right)\Big]} 
\Phiext
\equiv \gamma \Phiext ~.
\label{gamma}
\end{equation}
The different ingredients have the following physical origin.
The $J_3$ in the numerator represents the coupling of the magnetic field
to the internal gauge flux through the corresponding three-spin term.
On the other hand, the $J_4$ in the denominator represents the 
stiffness of the internal gauge field that originates from the 
four-spin exchanges.
For the \ET\ parameters, we estimate 
$g_3 J_3/(12 g_4 J_4 |\chi_0|) \simeq 0.7$.
The term in the square brackets in the denominator reflects the enhanced 
susceptibility to the fluxes discussed above; it can be also viewed as a 
suppression of the internal gauge field stiffness.
In particular, when this term goes to zero, the spinon Fermi sea state
becomes unstable to spontaneous flux generation even in the absence of 
the external field.
Given the proximity of the competing states, it is therefore reasonable 
to estimate this number to be of order one-half, so for the
\ET\ material we roughly estimate 
\begin{equation}
\gamma_{\kappa-{\rm (ET)}_2 {\rm Cu}_2 {\rm (CN)}_3} \;\simeq\; 
1-2 ~.
\end{equation}
Thus, we conclude that due to the triangular ring exchanges and the 
proximity to the flux (or antiferromagnetic) instability, 
the effective orbital field seen by the spinons is comparable and
can even be larger than the applied magnetic field!

\begin{figure}
\centerline{\includegraphics[width=\columnwidth]{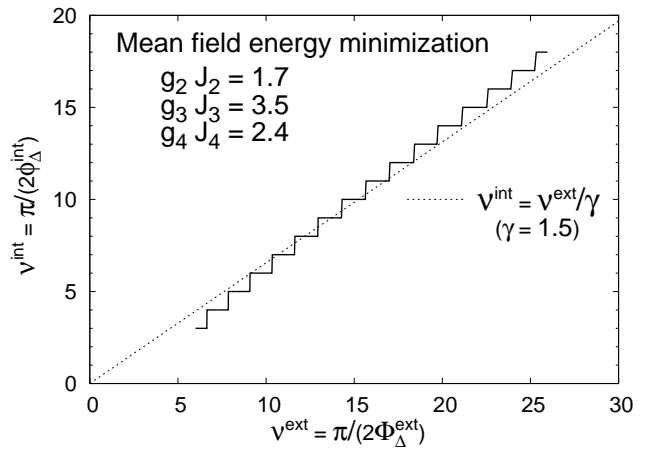}}
\vskip -2mm
\caption{
Result of the mean field energy minimization with respect to $\phiint$
in the presence of the magnetic field at zero temperature.
The optimal spinon filling factor $\nu^{\rm int}$ is plotted vs the 
nominal filling factor $\nu^{\rm ext}$ for the applied magnetic field.
The figure shows a sequence of first-order transitions with
$\nu^{\rm int}$ stepping through integer values.
The parameters roughly correspond to our estimates for the \ET\ 
from Sec.~\ref{sec:Hring}, 
with appropriate renormalization factors from Appendix~\ref{app:grenorm}.
The dotted line shows the overall trend Eq.~(\ref{gamma}).
The data plotted correspond to unrealistically large fields, 
but the same behavior is expected to continue for smaller laboratory 
fields.
}
\label{fig:nuopt}
\end{figure}

\vskip 1mm
We now examine the mean field energy Eq.~(\ref{emfphi}) more carefully.  
In the presence of the static internal gauge flux, the spinon spectrum 
consists of Landau bands.
For either spin species, the filling factor of these Landau levels is 
$\nu = \pi/(2 \phiint_\triangle)$, and the flux states are special 
when $\nu$ is integer.  Thus, the bond expectation value $|\chi_\phi|$ 
has upward-pointing cusps at these fluxes as can be seen in 
Fig.~\ref{fig:chiphi}.
The states with integer $\nu$ are therefore expected to be more stable.
In fact, when the mean field energy is minimized with respect to 
$\phiint$ using the material parameters quoted above, we find that the 
optimal such flux does not change continuously but instead goes through 
this discrete set corresponding to the integer filling of the spinon 
Landau levels.  This is shown in Fig.~\ref{fig:nuopt}.
The overall magnitude is still given by Eq.~(\ref{gamma}).

\vskip 1mm
Let us exhibit the structure of the discussed mean field in a more 
transparent way.  The reason behind the elementary manipulations 
below is to draw a connection with the so-called strong magnetic 
interaction effects in the magneto-oscillation studies in metals.
These are reviewed in Appendix~\ref{app:mint}, which contains useful
interpretations of the expressions below.
At the expense of being repetitious, what follows will allow us
to estimate appropriate effective parameters, which we will then use
in our discussion of the spinon system at finite temperature. 

First of all, we model the behavior of $|\chi_\phi|$,
Fig.~\ref{fig:chiphi}, as
\begin{equation}
|\chi_\phi| = |\chi_0|(1 + c\phi_\triangle^2) 
- \Xi^{\rm osc}_\phi ~,
\label{chiphi_model}
\end{equation}
where we separate the envelope Eq.~(\ref{chiphi_envelope}), 
which is the band structure effect, 
and the oscillating part $\Xi^{\rm osc}_\phi$ due to the 
discreteness of the spinon Landau levels.
The latter is well approximated by the familiar 
expression for the oscillating piece of the energy of the continuum 
Landau problem,
\begin{equation}
\Xi^{\rm osc}_\phi \propto \nu^{-2} (\nu - k)(k+1 - \nu)
\end{equation}
for the filling factor $\nu$ between integers $k$ and $k+1$.  
This approximation is also plotted in Fig.~\ref{fig:chiphi} with
a suitably chosen numerical amplitude for $\Xi^{\rm osc}_\phi$.
In particular, from the numerical data we estimate the cyclotron 
mass for the half-filled triangular lattice band to be 
$m_c \approx t_{\rm spinon}^{-1}$ in units with the lattice spacing 
set to one, while the corresponding cyclotron frequency at filling
$\nu$ is
\begin{equation}
\hbar\omega_c \approx 3.57 \, t_{\rm spinon}/\nu ~.
\label{omegac}
\end{equation}

Returning to the mean-field energy Eq.~(\ref{emfphi}), we expand to
quadratic order in $\phiint$ bearing in mind that the oscillating
piece $\Xi^{\rm osc}_\phi$ has an overall magnitude $(\phiint)^2$:
\begin{eqnarray}
\epsilon_{\rm mf} &=& \kappa_4
\left[(\phiint_\triangle)^2 
     - 2 \,\gamma\, \Phiext_\triangle \phiint_\triangle \right] 
+ \epsilon^{\rm osc} ~,
\label{emf_phys}
\\
\epsilon^{\rm osc} &=& \kappa_4\; y' \frac{\Xi^{\rm osc}_\phi}{|\chi_0|} 
\,=\, 12 \, t_{\rm spinon}\, \Xi^{\rm osc}_\phi ~,
\end{eqnarray}
with the effective parameters
\begin{eqnarray}
\kappa_4 &=& 192 g_4 J_4 |\chi_0|^4 (1-2cy) ~,
\label{kappa4}
\\
t_{\rm spinon} &=& 2 g_2 J_2 |\chi_0| + 32 g_4 J_4 |\chi_0|^3 ~.
\label{tspinon}
\end{eqnarray}
Here $\gamma$ is defined in Eq.~(\ref{gamma});
$y \equiv 1 + g_2 J_2 /(16 g_4 J_4 |\chi_0|^2)$;
and $y' \equiv 2y/(1-2cy)$.
We remind that the energy here is per triangular lattice site,
and see that $\kappa_4$ is the corresponding measure of the stiffness
of the internal gauge field---see Appendix~\ref{app:gt}.
We also see explicitly how the magnetic field acts to induce the 
internal gauge flux.
On the other hand, using Eq.~(\ref{chiphi_model}), we can interpret
$\epsilon^{\rm osc}$ as an oscillating piece of some fermion kinetic
energy of spinful fermions hopping on the triangular lattice with
amplitude $t_{\rm spinon}$.
It is useful to remember that we are analyzing the mean field energy 
Eq.~(\ref{emfphi}), and from this we are separating out what looks like 
an effective spinon kinetic energy.
The division of the mean field energy into the gauge field and the 
spinon parts is somewhat arbitrary, but represents a useful 
separation by the character of their dependence on $\phiint$.
In particular, we can now see the similarity with Eq.~(\ref{Omega_mf}) 
and thus connect with the studies of magnetic interaction effects in 
metals summarized in Appendix~\ref{app:mint}.
The above equations constitute the main result of this section.

To emphasize the variational energetics character of the mean field 
procedure and the intrinsically common origin of the gauge field
and the spinon parts, we do not use such suggestive separation 
explicitly in the treatment below.
A formal condition for an extremum of $\epsilon_{\rm mf}$ reads
\begin{eqnarray}
\gamma \Phiext_\triangle 
= \phiint_\triangle 
+ \frac{y'}{2|\chi_0|} \frac{\partial\Xi^{\rm osc}_\phi}
                            {\partial \phiint_\triangle}
\equiv \phiint_\triangle - {\EuFrak M}^{\rm osc}[\phiint_\triangle] ~,
\label{extrem}
\end{eqnarray}
which can be solved graphically by plotting the right-hand side
as a function of $\phiint$ and seeking crossings with the
horizontal line $\gamma\Phiext$.
This can be compared with Eq.~(\ref{Hext}) in Appendix~\ref{app:mint}.
Here we remark that the oscillations of the ``magnetization'' 
${\EuFrak M}^{\rm osc}$ have roughly the sawtooth pattern
familiar in two dimensions, with the amplitude which can
be estimated from the data in Fig.~\ref{fig:chiphi}.
Note that the internal gauge field stiffness is implicitly included in 
${\EuFrak M}^{\rm osc}$, cf.~Eq.~(\ref{Hext}).
We estimate that the corresponding $|{\EuFrak M}^{\rm osc}|_{\rm max}$ 
is as large as $1$. 
In particular, we conclude that at zero temperature essentially 
each oscillation period supplies such graphical solutions of 
Eq.~(\ref{extrem}).
Solutions are found even in the region with $\nu \sim 1$
(corresponding to very large internal flux) even when $\Phiext$
is small.  Remembering the cusps in $|\chi_\phi|$, this means
that each integer filling gives a locally stable minimum, which
we can also verify directly by plotting $\epsilon_{\rm mf}$ as
a function of $\phiint$.  The global minimum is given by an
integer filling near that corresponding to Eq.~(\ref{gamma}).
This provides a more complete description of our direct minimization 
result Fig.~\ref{fig:nuopt} at zero temperature, and is useful when 
generalizing to finite temperature.

The multiple solutions signify an 
\emph{instability of the flux states with continuously varying 
$\phiint$}.  The instability occurs whenever the derivative of the 
right-hand side of Eq.~(\ref{extrem}) becomes negative:
\begin{equation}
\frac{\partial {\EuFrak M}^{\rm osc} }{\partial \phiint_\triangle} > 1
\;\; \implies \;\;
\nu^2 \, |{\EuFrak M}^{\rm osc}|_{\rm max} \gtrsim 1 ~.
\label{instability}
\end{equation}
This is just a more formal restatement of our discussion in the 
preceding paragraph:
given $|{\EuFrak M}^{\rm osc}|_{\rm max} \sim 1$,
the instability condition is satisfied all the way to
very large internal fields corresponding to $\nu \sim 1$.
This is unlike the ordinary metals where the largest magnetic
field that allows such an instability is typically much smaller.
As explained in Appendix~\ref{app:mint}, the origin of this
is the very small internal gauge field stiffness, which
is implicitly built into the numerical estimate
$|{\EuFrak M}^{\rm osc}|_{\rm max} \sim 1$ in the above formalism.
However, such large numerical value of 
$|{\EuFrak M}^{\rm osc}|_{\rm max}$ is unavoidable since it is 
the same mean field energetics that stabilizes the Fermi sea state
and determines both the gauge field stiffness and the spinon 
kinetic energy.

\subsection{Mean field description at finite temperature}
The preceding analysis was performed at zero temperature.
Also, we ignored the Zeeman spin coupling to the magnetic field.  
We first note that for integer filling $\nu$ the spinon spectrum has a 
gap given roughly by the corresponding cyclotron frequency.
Such an integer quantum Hall state of spinons is in fact a chiral
spin liquid\cite{chiralSL} induced by the magnetic field, and the above 
treatment suggests a sequence of first-order transitions stepping
through such states rather than a continuous variation of $\phiint$
(see Fig.~\ref{fig:nuopt}).  
Finite temperature smears the effects of the discreteness of the 
Landau levels, and the instability of the flux states with continuously 
varying $\phiint$ becomes weaker.

To proceed more quantitatively, we again use mean field and apply 
standard magneto-oscillation results at finite-temperature in two 
dimensions supplemented with our estimates of the effective parameters 
such as the cyclotron frequency Eq.~(\ref{omegac}).
Thus, we can roughly incorporate the effects of the finite temperature 
and the Zeeman spin splitting by multiplying the oscillating 
magnetization ${\EuFrak M}^{\rm osc}$ by the corresponding 
suppression factors $R_T$ and $R_S$ from Eq.~(\ref{R_T}) in 
Appendix~\ref{app:mint}.
The instability condition Eq.~(\ref{instability}) then reads
\begin{equation}
R_T R_S \; \nu^2 \, |{\EuFrak M}^{\rm osc}|_{\max} \gtrsim 1 ~.
\end{equation}
For a fixed external magnetic field, we can use Eq.~(\ref{gamma})
to estimate the corresponding spinon filling $\nu$ in the internal 
gauge field.  Focusing on the temperature suppression factor,
the homogeneous flux state becomes unstable when the temperature is below
\begin{equation}
T_{\rm instab.} = \frac{x \hbar\omega_c}{2\pi^2} ~,
\quad\quad
\frac{\sinh(x)}{x} \approx \nu^2 |{\EuFrak M}^{\rm osc}|_{\rm max} ~.
\label{Tinstab}
\end{equation}
For high filling factors $\nu$ and using
$|{\EuFrak M}^{\rm osc}|_{\rm max} \sim 1$, we estimate with logarithmic
accuracy $x \sim 2\log(\nu)$.
We give numerical estimates for the \ET\ material in Sec.~\ref{sec:ET}.

Let us briefly think what happens beyond the mean field.
The chiral spin liquid states are stable topologically ordered 
phases\cite{chiralSL} at zero temperature, but strictly speaking 
do not survive on long length scales at any nonzero temperature in 
two dimensions.
Instead, the spin system can be continuously connected to the 
featureless high-temperature paramagnet.
In this case, we interpret the mean field estimate Eq.~(\ref{Tinstab})
as delineating the temperature-field regime in which the spin system 
becomes particularly ``soft'' in its response to the magnetic field.  
It is also possible for a $T=0$ first-order transition between
chiral states to extend to small finite temperatures even though the 
paramagnetic phases on both sides do not have a symmetry or 
topological distinction anymore.
Note also that in the above schematic treatment, we have not 
considered the temperature dependence of the effective parameters
For example, in the formal mean field, the effective $\tspinon$ 
decreases with temperature and disappears above $T \sim 50-100$K
for the \ET\ parameters (this can be viewed as a crude estimate
of the temperature below which quantum spin correlations develop).
Such detailed questions on the behavior of the spin system at finite 
temperature remain open for future investigations.

\subsection{Inhomogeneous state in the physical system at low 
temperatures}
\label{subsec:physset}
We conclude the mean field description with a discussion of
some experimentally relevant aspects.
Our analysis indicates the instability of the homogeneous flux states
with continuously varying $\phiint$ at temperatures below 
$T_{\rm instab.}$.
Throughout, we treat the external magnetic field as fixed,
and the instability represents a strong nonlinear back action of the 
spinons onto the internal gauge field in the presence of such
fixed external source.
Treating the magnetic field as fixed is justified since, 
as we describe in Appendixes~\ref{app:gt}~and~\ref{app:mint}, 
the electromagnetic field is six to seven orders of magnitude more stiff 
than the internal gauge field and will not adjust itself to the 
electronic system until much lower temperatures.
Of course, there is a slight effect on the local magnetic field
depending on the state of the spin system, but for example
for the uniform flux mean field state we estimate the orbital
contribution to be one order of magnitude smaller than the
Pauli spin contribution.\cite{estimate_chiorb}
The physical setting therefore has fixed $\Phiext$.
In particular, we do not have access to different sample 
``demagnetization geometries'' considered for the magnetic interaction 
phenomenon in conventional metals\cite{Shoenberg, Abrikosov}
-- the internal gauge field does not 
``leak out'' of the sample.

In such setting in an ideal crystal system at zero temperature, 
we predict a sequence of first-order transitions,\cite{true_inhom} 
but each phase is still a uniform flux state with the corresponding 
integer $\nu$.
This situation, however, is highly susceptible to large-scale 
imperfections of a real system, and an outcome with many domains
is likely.
Another important consideration is the possible crystal mosaic 
in the sample, since the discussed orbital effects are determined
 by the component of the magnetic field $H_\perp$ normal to the 
two-dimensional plane.
A detailed characterization of the possible inhomogeneities clearly
requires much more material knowledge.

\section{Application to \ET}
\label{sec:ET}
Throughout, we have used the \ET\ parameters to motivate various
approximations and make the otherwise formal discussion more physical.
Here we want to focus more on the material itself and collect the 
relevant numbers in one place.

We first want to point out that the material is rather clean.
Thus, we estimate $k_F l \gtrsim 50$ in the metallic phase at $0.8~$GPa.
Shubnikov-deHaas (SdH) oscillations are observed\cite{ET_SdH} at 
temperatures around $1.5~$K, which is another indication of the 
material quality.  The SdH signal is consistent with the Fermi surface
and the one-band model description.
For reference, the Landau level filling factor for the \ET\ material
is given by $\nu^{\rm ext} = 3595/H[{\rm Tesla}]$.
In Appendix~\ref{app:mint}, we estimate the characteristic magnetic field
$H_0 \approx 5~$T and temperature $T_{\rm dm} \approx 0.16~$K for the 
magnetic interaction instability\cite{DiamagPT, Condon,reviewCondon} 
in this metallic phase coupled to the electromagnetic field.

Turning to the spin liquid phase, in Ref.~\onlinecite{ringxch1} 
we used the low-temperature susceptibility to estimate the spinon 
hopping amplitude $t_{\rm spinon} \simeq 350~$K.
We can also use the mean field estimate Eq.~(\ref{tspinon}),
which gives a smaller but reasonable value 
$t_{\rm spinon} \simeq 100~$K.
Using Eqs.~(\ref{omegac}) and (\ref{Tinstab}), and 
$\nu^{\rm int}=\nu^{\rm ext}/\gamma$ with $\gamma \simeq 1-2$, 
we obtain the characteristic temperature at $B=8~$T to be 
$T_{\rm instab.} \simeq 1-2~$K.
Above this temperature, we expect uniform flux 
state.\cite{estimate_chiorb}

We now summarize qualitative predictions for the inhomogeneous state 
induced by large magnetic fields at low temperatures below
$T_{\rm instab}$.
As discussed in the paragraph following Eq.~(\ref{extrem}), 
the characteristic field $H_0$ for the spinon
system is beyond any practical fields, and we predict that in the
laboratory fields the instability temperature will increase roughly
linearly with the applied field, Eq.~(\ref{Tinstab}).  
Manifestations of this instability become more pronounced with 
increasing field and decreasing temperature.
Since this is an orbital effect, we expect distinction 
for fields perpendicular and parallel to the conducting planes.
One consequence of this scenario is that magneto-oscillation
measurements in their usual sense cannot be used to detect the
spinon Fermi surface in the spin liquid state considered here.
This is because the regime where such oscillations can become
visible is likely preempted by the instability of the homogeneous state.
Since the instability is in some sense an extreme manifestation of the
magneto-oscillations, conditions of the magnetic field uniformity
and the crystal mosaic are important experimental considerations,
as discussed in Sec.~\ref{subsec:physset}.
Other measurements in the spin liquid phase performed in high magnetic 
field and at low temperature may also require more careful 
interpretation.
Finally, some different direct probes of the Fermi surface such as
magneto-acoustic resonance that are less sensitive to the field
homogeneity may still be possible.

\section{Conclusions}
The main content of this work was summarized in the introduction,
and possible application to the \ET\ experiments including
some numerical estimates was discussed in the preceding section.
Here we want to conclude with a more general point.
While there has been a significant progress in the theoretical
understanding of exotic quantum phases, the lack of the material
evidence is viewed as a major obstacle.  Recently, several candidate 
spin liquid systems have been found in frustrated spin 
systems,\cite{ET_SL1, CsCuCl, Kagome} of which the \ET\ 
is a very promising example.  This material motivated
and guided the detailed theoretical considerations in the
present work on the magnetic field response of the spin liquid state 
stabilized near the Mott transition.  
Without such a guide, the proposed effects would be hard to anticipate.
More work matching experiment and theory focusing on material
properties will likely be fruitful in developing our understanding
and in further pursuits of such unusual quantum phases.

One specific motivation for the present study has been the
anomalous NMR line broadening reported in 
Refs.~\onlinecite{Kanoda_kitp}~and~\onlinecite{Kawamoto}.
While some of our expectations for the inhomogeneous state
at low temperatures resemble the experimental phenomenology,
the gradual development of the broadening starting from rather
high temperatures and also the magnitude of the inhomogeneous fields
are more suggestive of a stronger impurity mechanism.
Experiments in parallel field can further clarify the role of
the impurity and orbital effects.
The main predictions in the present paper of the spinon orbital
field and the fragility of the spinon Fermi sea state in the
applied magnetic field represent intrinsic response of the spin liquid
and are expected to play an important role at low temperatures.
It is hoped that our proposals will stimulate further experimental
and theoretical questions.

{\it Note added:}
Recently, the author learned of Ref.~\onlinecite{SenChitra}, 
which also discusses the coupling of the external field to the spin 
chirality and considers possible effects in some different contexts 
for Hubbard systems with triangles.

\acknowledgments
The author has benefited from useful discussions with M.~P.~A.~Fisher, 
V.~Galitski, Y.-B.~Kim, P.~A.~Lee, S.~S.~Lee, A.~Paramekanti, 
T.~Senthil, and A.~Vishwanath, and also thanks the 
Aspen Center for Physics Summer Program on 
``Gauge Theories in Condensed Matter Physics'' 
where part of this work was completed.
The research at KITP is supported through NSF grant PHY-9907949.
Use of Hewlett-Packard and CNSI Computer Facilities at UCSB
is acknowledged.

\appendix

\section{Hubbard to Heisenberg in the presence of magnetic field}
\label{app:Hring}
This appendix gives the effective spin Hamiltonian to order $t^4/U^3$ 
for a general lattice Hubbard system with complex hopping 
$t_{rr'} = |t_{rr'}| e^{i A_{rr'}}$, which extends previously available 
results\cite{t/U} to the case in the presence of the magnetic field.

The starting point is the Hubbard Hamiltonian at half-filling
\begin{equation}
H = U \sum_r n_{r\up} n_{r\dn}
- \sum_{rr'} t_{rr'} c_{r\sigma}^\dagger c_{r'\sigma} ~,
\end{equation}
with $t_{r'r}=t_{rr'}^*$.  The hopping part is treated as a perturbation,
and a canonical transformation is performed into the sector with no 
double occupancy.  The effective Hamiltonian to order $t^4/U^3$ reads
\begin{eqnarray}
H_{\rm eff} &=& H_2 + H_3 + H_4 ~, \\
H_2 &=& \sum_\bondpic \frac{2 |t_{12}|^2}{U} (P_{12} - 1) ~, \\
H_3 &=& \sum_\looptripic \frac{6 {\rm Im}(t_{12} t_{23} t_{31})}{U^2}
\,i\, (P_{123} - \Hc) ~,
\end{eqnarray}
\begin{widetext}
\begin{eqnarray}
H_4 &=& 
\sum_\bondpic \frac{8 |t_{12}|^4}{U^3} (1-P_{12})
+ {\sum_{1,2,3}}^\prime \frac{|t_{12}|^2 |t_{13}|^2}{U^3} (P_{23} - 1) \\
&+& \sum_\loopfourpic
\frac{{\rm Re}(t_{12} t_{23} t_{34} t_{41})}{U^3} 
\Big[ 20 (P_{1234} + \Hc)
        - 12 (P_{12} + P_{23} + P_{34} + P_{41} + P_{13} + P_{24}) + 32
\Big] ~.
\end{eqnarray}
\end{widetext}
Here $P_{12}$, $P_{123}$, and $P_{1234}$ denote two-spin, three-spin, and
four-spin exchange operators respectively.  The latter two operators
move the spins in a ringlike manner.  For example, the three-spin 
exchange in terms of the fermions is
$P_{123} = (c_{1\alpha}^\dagger c_{1\beta})
(c_{2\beta}^\dagger c_{2\gamma}) (c_{3\gamma}^\dagger c_{3\alpha})$,
and acts on the spins as 
$P_{123} : |\sigma_1, \sigma_2, \sigma_3\ra 
\to |\sigma_3, \sigma_1, \sigma_2 \ra$.
To simplify the final expressions, we used the identities 
Eqs.~(\ref{ReP3},\ref{ImP4}) specific to the spin-1/2 case.

The sum in $H_2$ is over all bonds of the lattice.  The sum in $H_3$
is over all three-site loops, where each group of three sites connected 
by the links enters the sum precisely once.  Similarly, the sum
in the second line for $H_4$ is over four-site loops, which are 
counted with no base point or orientation.
Finally, the primed sum in the first line for $H_4$ is over distinct
sites $1,2,3$ with nonzero $t_{12}$ and $t_{13}$ (each contribution
will therefore appear two times).
Eq.~(\ref{Heff}) in the main text is obtained by specializing 
to the isotropic triangular lattice.

Note that the complex phase of a loop product of $t_{rr'}$ measures 
the flux of the magnetic field through the loop.
Without such fluxes, $H_3$ vanishes and after some transformations 
we reproduce the result of Ref.~\onlinecite{t/U}.

As a side remark, we observe that in the absence of three-site
loops, there is no linear coupling to the magnetic field to this
order in $t/U$.  One can also show that for a half-filled Hubbard model
on a bipartite lattice, the effective Hamiltonian in the presence of 
the magnetic field can only contain terms that are even in the
applied field.

\vskip 2mm
Finally, it is also useful to give an expression for an effective
local electrical current operator.
In the original Hubbard model, the current on a link $1 \to 2$ is 
$I_{1 \to 2} = 
\frac{e}{i\hbar} (t_{12} c_{1\sigma}^\dagger c_{2\sigma} - \Hc)$.
Upon the canonical transformation to the lowest nontrivial 
order in $t/U$, the current operator becomes in terms of the 
spin variables
\begin{equation}
I^{\rm eff}_{1\to 2} = \frac{e}{\hbar} \sum_3 
\frac{6 {\rm Re}(t_{12} t_{23} t_{31})}{U^2} \,i\, (P_{123} - \Hc) ~.
\label{Ieff}
\end{equation}
We can use this expression e.g.~to estimate the physical magnetic field 
produced by the underlying orbital motion in the chiral spin states 
discussed in the main text.\cite{estimate_chiorb}

\section{Renormalized mean field}
\label{app:grenorm}
\begin{figure}
\centerline{\includegraphics[width=\columnwidth]{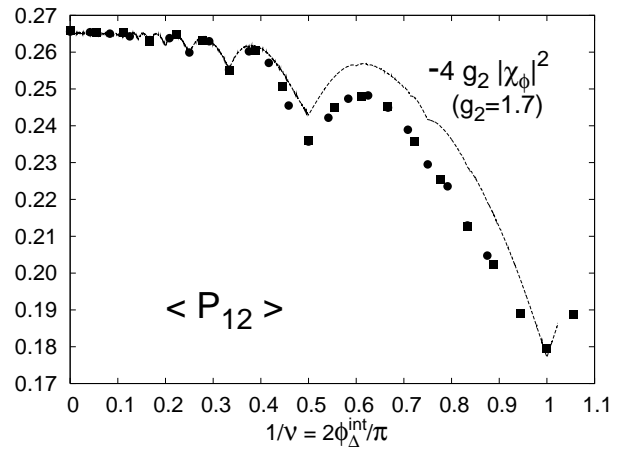}}
\vskip -2mm
\caption{Expectation value of the two-spin exchange $P_{12}$
on each link evaluated in the Gutzwiller-projected uniform flux state.
The data is for commensurate fluxes on triangular lattice cuts
with sizes $36\times 12$ (squares) and $48 \times 12$ (circles) 
with periodic boundary conditions.
The line shows the renormalized mean field estimate with $g_2 = 1.7$ 
(the mean field data is obtained using Fig.~\ref{fig:chiphi}).  
Since we are interested only in the variation with $\phiint$, 
a constant offset is added to the mean field values.
}
\label{fig:P12}
\end{figure}

\begin{figure}
\centerline{\includegraphics[width=\columnwidth]{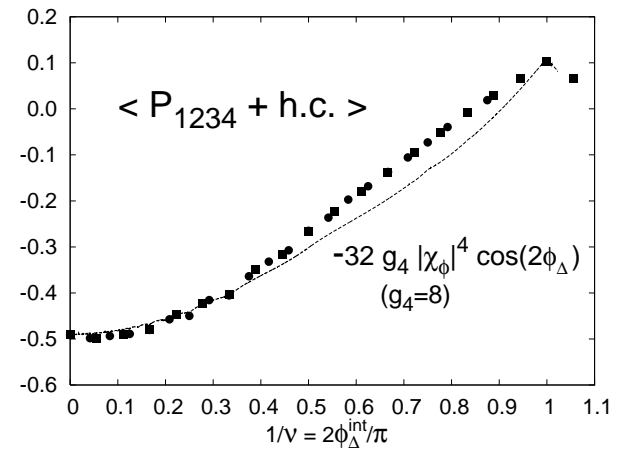}}
\vskip -2mm
\caption{
Same as in Fig.~\ref{fig:P12}, but for the ring exchange 
operator $P_{1234} + \Hc$ around a rhombus.
}
\label{fig:rP1234}
\end{figure}

\begin{figure}
\centerline{\includegraphics[width=\columnwidth]{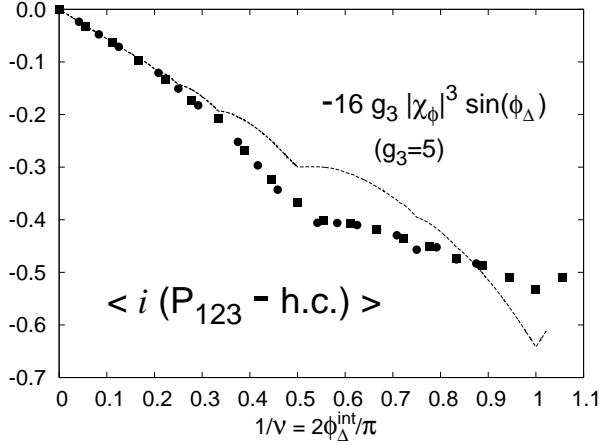}}
\vskip -2mm
\caption{
Same as in Fig.~\ref{fig:P12}, but for the three-spin operator
$i(P_{123} - \Hc)$ around an elementary triangle.
The orientation of the triangular loop $1 \to 2 \to 3 \to 1$
coincides with that of $\phiint_{123} = \phiint_\triangle$.
}
\label{fig:iP123}
\end{figure}

In this appendix, we estimate the renormalization factors
$g_2$, $g_4$, and $g_3$ used in the mean field treatment in
Sec.~\ref{sec:mf}.  We consider the uniform flux states and 
measure the relevant expectation values of $P_{12}$,  $(P_{1234}+\Hc)$, 
and $i(P_{123}-\Hc)$ in the Gutzwiller-projected wavefunctions.
The variation of the expectation values with $\phiint$ is matched
with the corresponding mean field estimates---see Eq.~(\ref{genEmf}).
Figures~\ref{fig:P12},~\ref{fig:rP1234},~and~\ref{fig:iP123} 
show this matching and are self-explanatory.
We note that the Gutzwiller wavefunction evaluations can be
performed only for small system sizes and large values of $\phiint$.
However, we expect the same trend to remain for small fluxes as well.
For small fluxes, an analytical mean field treatment is then
performed (Sec.~\ref{sec:mf}) using the estimated renormalization 
factors.  This is the main idea of the so-called renormalized 
mean field approach.\cite{renormmf}

\section{Spinon-gauge theory}
\label{app:gt}
In the main text, we often use the gauge theory language,\cite{LeeNagaosaWen}
and it is useful to state the connection more explicitly.  
We follow the treatment of Ref.~\onlinecite{SSLee} as an example.
One starts with a slave-rotor representation
$c_{r\sigma}^\dagger = f_{r\sigma}^\dagger e^{i\theta_r}$
of the Hubbard model and obtains the following effective theory for the 
spinons $f_{r\sigma}$ and chargons $e^{i\theta_r}$:
\begin{eqnarray}
\label{Sgt}
S &=& \int d\tau \left({\cal L}_f + {\cal L}_\theta \right) ~,\\
{\cal L}_f &=& 
\sum_r f_{r\sigma}^\dagger (\partial_\tau - i a_r^0 - \mu) f_{r\sigma}
- \sum_{rr'} t_f e^{i a_{rr'}} f_{r\sigma}^\dagger f_{r'\sigma} ~, 
\nonumber \\
{\cal L}_\theta &=& 
\frac{1}{2U} \sum_r (\partial_\tau \theta_r - a_r^0)^2
- \sum_{rr'} t_\theta 
e^{i (\theta_r - \theta_{r'} - a_{rr'} + A^{\rm ext}_{rr'})} ~.
\nonumber
\end{eqnarray}
The imaginary time is continuous, while the space lattice is retained 
to indicate similarity of the spinon part with that in
Eq.~(\ref{Htrial}).
The spinons and the chargons are coupled with opposite charge to the
internal gauge field $({\bf a}, a^0)$, whose spatial components 
represent fluctuations of the phase of the spinon hopping field, 
$\tspinon_{rr'} = t_f e^{i a_{rr'}}$, 
while the temporal $a^0_r$ implement the slave-particle constraints.  
The coupling to the external field $A^{\rm ext}$ is included 
schematically by assigning the electron charge to the chargon field.

Assuming the chargons are gapped and integrating them out, 
we are left with the spinon-gauge system where the gauge field now
has some stiffness $\kappa$,
\begin{equation}
\label{La}
{\cal L}_a = \frac{\kappa}{2} 
({\bm \nabla} \wedge {\bf a} - {\bm \nabla} \wedge {\bf A}^{\rm ext})^2
+ \frac{\kappa_\tau}{2} (\partial_\tau {\bf a} - {\bm \nabla} a^0)^2 ~.
\end{equation}
Here 
${\bm \nabla} \wedge {\bf a} = \partial_x a_y - \partial_y a_x \equiv b$;
we separated this ``spatial'' part for later convenience.

Our mean field treatment Eq.~(\ref{genEmf}) roughly corresponds to
considering the spinon-gauge system with static but possibly 
spatially varying gauge field ${\bf a}$.  
Observe that the internal gauge field obtains its stiffness after
integrating out the massive chargons, and also observe how the 
magnetic field couples to the internal gauge field.
This is similar to our discussion of the effective spin Hamiltonian,
where the crucial ingredient is also charge fluctuations.  
Note however that the effective Hamiltonian treatment is more explicit. 
Thus, the gauge field stiffness originates primarily from the 
four-spin ring exchanges, while the coupling to the magnetic field 
comes from the three-spin processes -- see 
Eqs.~(\ref{genEmf})~and~(\ref{emf_phys}).  
In the schematic gauge theory derivation this distinction is not
being made.  The result $\gamma=1$ of such schematic treatment
roughly agrees with our estimate using Eq.~(\ref{gamma}), but this
is rather fortuitous -- e.g., as we mention towards
the end of Appendix~\ref{app:Hring}, in a Hubbard model on
a lattice with no elementary triangles the linear coupling of the
internal gauge flux to the applied magnetic field does not appear
even to order $t^4/U^3$.
To capture the detail present in our Eq.~(\ref{gamma}) in the gauge 
theory approach, one needs to go back to the original slave particle
rewriting and carefully consider saddle point conditions and integrations
over massive fields when deriving Eq.~(\ref{Sgt}) in the presence of
the external field.
As an example, the suppression of the gauge field stiffness due to 
the proximity to the flux phase discussed after Eq.~(\ref{gamma}) 
originates from some modes which become soft near the flux phase.
The effective spin Hamiltonian approach is more transparent in this
respect and also allows quantitative estimates of the physical 
quantities.

In particular, from Eq.~(\ref{kappa4}) we quote the internal gauge 
field stiffness in terms of the energy cost per triangular lattice site,
$\epsilon_{\rm int} = \kappa_4 (\phiint_\triangle)^2$,
\begin{equation}
\kappa_4 \;\sim\; 192 g_4 J_4 |\chi_0|^4 
\;\sim\; 0.14\, g_4 J_4 
\;\sim\; 2 ~{\rm meV} ~.
\end{equation}
Here we used the Sec.~\ref{sec:Hring} estimate $J_4 \simeq 2$~meV,
and used the Fig.~\ref{fig:rP1234} estimate $g_4 \simeq 8$.

This can be compared with the corresponding stiffness of the
physical electromagnetic (EM) field in the bulk of the \ET\ compound,
$\epsilon_{\rm EM} = V B^2/(8\pi) \equiv 
\kappa_{\rm EM}(\Phiext_\triangle)^2$,
\begin{equation}
\kappa_{\rm EM} = \frac{V \hbar^2 c^2}{8 \pi A_\triangle^2 e^2}
= 1.1\times 10^4 ~{\rm eV} ~.
\end{equation}
Here $V=850.6~$\AA$^3$ is the 3d volume per triangular lattice site, 
$A_\triangle = 28.76~$\AA$^2$ is the area of an elementary triangle,
and $\Phiext_\triangle = e B A_\triangle/(\hbar c)$ is the
appropriate dimensionless external flux.

We therefore conclude that the bare internal gauge field stiffness
is $\sim 10^7$ times smaller than the electromagnetic field stiffness.
As we discuss in the following Appendix~\ref{app:mint},
this makes a dramatic difference in the response of the spinon-gauge 
system to the external field compared with that of electrons in a 
conventional metal.

\section{Analogy with magnetic interaction effects in metals}
\label{app:mint}
The purpose of this appendix is to clarify the status of our mean field
treatment in Sec.~\ref{sec:mf} by pointing out the analogy with
the so-called  magnetic interaction effects in the study of 
magneto-oscillations in metals.
This material is available in textbooks.\cite{Shoenberg, Abrikosov}
The presentation below parallels some of our discussion in
Sec.~\ref{sec:mf} in a different language, but we hope
that the reader will benefit from this duplication.

For ease of reference, we first write down the oscillating part 
of the magnetization of a two-dimensional electron gas in a field 
$H$ at a finite temperature $T$:
\begin{equation}
{\cal M}^{\rm osc}(H) = 
R_T R_S \frac{n e\hbar}{\pi m^* c} 
\sin\left(\frac{2\pi F}{H} \right) ~.
\label{Mosc}
\end{equation}
Only the main harmonic is shown.
The magnetization is given per unit area, and $n$ is the 2d electron 
density (including spin); $m^*$ is the cyclotron mass; 
$F = n \pi c\hbar/e$ so that $F/H = \nu$ is the Landau level filling 
for each spin species.
The suppression factors $R_T$ and $R_S$ due to finite
temperature and Zeeman spin splitting are
\begin{equation}
R_T = \frac{2\pi^2 T/\hbar\omega_c}{\sinh(2\pi^2 T/\hbar\omega_c)}~;
\quad\quad
R_S = \cos\frac{g \pi m^*}{2 m_e} ~,
\label{R_T}
\end{equation}
where $g$ is the spin g-factor.

The essence of the magnetic interaction effects is the back action of the
electrons onto the electromagnetic field, which can be significantly 
enhanced by the oscillating character of ${\cal M}$.
Specifically, consider fermions coupled to a dynamical gauge field $a$ 
in the presence of the external field $A^{\rm ext}$, with the 
Lagrangian density ${\cal L}_f + {\cal L}_a$ given by 
Eqs.~(\ref{Sgt},\ref{La}). 
The microscopic magnetic field is $b = \nabla \times a$ 
and the external field is $H^{\rm ext} = \nabla \times A^{\rm ext}$.
In the case of electrons, $a$ represents the physical electromagnetic 
field and $\kappa$ is the appropriate EM stiffness;
e.g., in a 2d layered material the magnetic field energy per unit area is
$d^* b^2/8\pi$ where $d^*$ is the spacing between the layers.

The key observation is that electrons see the average microscopic field 
$B$ because the typical Larmor orbits are large.
The mean field treatment is then to assume a static but possibly
spatially varying such field $B(r)$ and solve the Landau problem for 
electrons moving in this field.  The mean field functional including the 
energy of the gauge field is\cite{Shoenberg, Abrikosov, Holstein}
\begin{equation}
\Omega_{\rm mf} = \Omega_{\rm ferm}(B) 
+ \frac{\kappa}{2}(B-H^{\rm ext})^2 ~,
\label{Omega_mf}
\end{equation}
where $\Omega_{\rm ferm}(B)$ is the free energy density for fermions
in the static field $B(r)$.  
The mean field functional is to be minimized with respect to $B(r)$.  
If we look for a uniform solution, we obtain 
\begin{equation}
H^{\rm ext} = B 
+ \frac{1}{\kappa} \frac{\partial \Omega_{\rm ferm}}{\partial B}
\equiv B - \frac{1}{\kappa} {\cal M}(B) ~,
\label{Hext}
\end{equation}
where in the last equation we defined the magnetization density
${\cal M}(B)$.  For a 2d electron gas, the oscillating piece of the 
magnetization is given in Eq.~(\ref{Mosc}), and the mean field treatment 
is to replace $H \to B$.

Eq.~(\ref{Hext}) is to be solved for $B$.  It is customary to plot
the right hand side as a function of $B$ and seek the intersection
with the horizontal line $H^{\rm ext}$.  When this is unique, 
the corresponding uniform field is the sought for stable solution.  
However, ${\cal M}(B)$ contains an oscillating piece, and if
we find that
\begin{equation}
\frac{1}{\kappa} \frac{\partial \cal M}{\partial B} > 1 ~,
\end{equation}
then the solution is no longer unique, which signals an instability.
Note that while the amplitude of the oscillations
$|{\cal M}|_{\rm max}(T=0) = n\, e\hbar/(2m^* c)$
is small in metals compared to typical $\kappa H$, 
the amplitude of the ``susceptibility'' $\partial{\cal M}/\partial B$ 
obtains an additional factor $\sim F/H^2$, and the instability condition
can be satisfied at low temperatures and not too large fields.
After simple transformations, the instability condition reads
\begin{equation}
R_T R_S \; n\; \frac{\hbar^2 k_F^2}{2 m^*} 
\;\; \gtrsim \;\; \frac{\kappa}{2} H^2 ~,
\end{equation}
where $k_F$ is the Fermi wavevector.
When this condition is satisfied, the gain in the fermion energy
when the field $B$ is adjusted to obtain integer Landau level filling
overweights the magnetic field energy cost.
Since $R_T$ and $R_S$ do not exceed $1$, the above equation sets the 
maximal value $H_0$ for the instability to occur at zero temperature.  
On the other hand, at a finite temperature, we also require the Landau 
levels to be resolved, which is determined by $R_T$.
Roughly, for fields of order $H_0$, we require 
$2\pi^2 T \lesssim \hbar \omega_c$ with $\omega_c$ set by $H_0$.

The 2d Landau problem for free electrons can be solved beyond the above 
single-harmonic treatment, and Ref.~\onlinecite{DiamagPT} contains such 
expressions for the phase boundary of the instability region.  
For example, the field $H_0$ is determined from
\begin{equation}
n\; \frac{\hbar^2 k_F^2}{2 m^*} = \kappa H_0^2 = d^* \frac{H_0^2}{8\pi}~.
\end{equation}
Ref.~\onlinecite{DiamagPT} also contains formulae directly appropriate 
for the layered organic materials.  
As a numerical application, consider the metallic phase of the \ET\ 
compound obtained under pressure of $0.76~$GPa, in which 
magneto-oscillations were reported in Ref.~\onlinecite{ET_SdH}.
Using $\epsilon_F \approx 100$meV, the cyclotron mass ratio
$\eta_c = m^*/m_e \approx 4$, the inter-layer spacing $d^*=14.84~$\AA,
and ignoring sample demagnetization effects,
we estimate $H_0 \approx 5~$T;
the peak temperature on the phase boundary of the instability regime
is estimated to be $T_{\rm dm} \approx 0.16$K.

Various aspects of what happens once the homogeneous state becomes
unstable are discussed in textbooks.\cite{Shoenberg, Abrikosov}  
There is also a growing recent literature on the observation of 
Condon domains\cite{Condon} in metals,
see Ref.~\onlinecite{reviewCondon} and references therein.

\vskip 1mm
The point that we want to make is that our mean field treatment
in Sec.~\ref{sec:mf} has the same character as the described 
treatment of the magnetic interaction effects in metals.
Thus, our mean field energy Eq.~(\ref{emf_phys}) can be readily 
identified with Eq.~(\ref{Omega_mf}), while the self-consistency
condition Eq.~(\ref{extrem}) corresponds to Eq.~(\ref{Hext}).
Clearly, the treatment of the magnetic interaction effects requires 
taking into account the energy of the gauge field.
This is implicit in our mean field treatment of Sec.~\ref{sec:mf}.
At this mean field stage, we do not need to disentangle the gauge field 
energy from the total energy and worry about the emergent nature of the 
gauge field, and this makes the procedure more simple.

An important difference between the spinon-gauge system and
the electrons in metals coupled to the EM field is the
very small gauge field stiffness in the spinon case,
as we estimated in Appendix~\ref{app:gt}.  
Because of this, the instability condition is satisfied more 
readily in the spinon-gauge system.  Thus, because the internal
gauge field stiffness is $10^6-10^7$ times smaller while the kinetic
parameters of spinons are not far from those of electrons, the above 
estimate of $H_0$ is to be multiplied by a factor of order $10^3$.
This is an impractically large field, and the spinon-gauge system
in the laboratory is in the regime of much smaller fields.
In this regime, the temperature at which the instability occurs
in a given field $H$ can be estimated as
\begin{equation}
T_{\rm instab.} = \frac{x \hbar\omega_c}{2\pi^2} ~,
\quad\quad
\frac{\sinh(x)}{x} \approx 
\frac{n\; \hbar^2 k_F^2/m^*}{\kappa H^2} ~.                     
\end{equation}
This temperature depends on the stiffness only logarithmically,
but as numerical estimates in Sec.~\ref{sec:mf} show, even under 
the logarithm the seven orders of magnitude difference in the
relevant stiffnesses produces a much wider temperature range for the
instability in the spinon-gauge system.



\begin{thebibliography}{10}
\bibitem{DiD}
Other chiral spin liquid states such as a Gutzwiller-projected $d+id$ 
state can also improve the trial energy in the presence of the magnetic 
field.  The present work focuses on the flux states for their simplicity,
but the detailed energetics including other competing states has 
not been explored.
\bibitem{estimate_chiorb}
As a crude estimate of the effect on the electromagnetic field,
we can calculate the orbital magnetic moment of an elementary
triangle to be
$\la m_{\triangle}^{\rm orb} \ra = \frac{e A_{\triangle}}{\hbar c} J_3
\la i (P_{123} - \Hc) \ra$,
where we used the expression Eq.~(\ref{Ieff}) for the electrical current
operator.  Using the mean field estimates 
Eqs.~(\ref{genEmf}, \ref{gamma}) (see also Fig.~\ref{fig:iP123}),
we evaluate that in the field of $8~$Tesla such orbital magnetic 
moment per site is of the order of $10^{-4} \mu_B$.  
For comparison, this applied field induces spin magnetic moment 
roughly equal to $3 \cdot 10^{-3} \mu_B$ per site, 
which is more than an order of magnitude larger;
to make the latter estimate we used the experimental low-temperature 
susceptibility of $\chi = 2 \cdot 10^{-4}~$emu/mol from
Refs.~\onlinecite{ET_SL1}-\onlinecite{Kanoda_kitp}.
\bibitem{true_inhom}
An interesting possibility discussed in the literature
(see e.g.~Ref.~\onlinecite{Abrikosov} and references therein)
is a ``true'' inhomogeneous flux state that would be inhomogeneous
also in an ideal crystal and in any geometry.
The analysis predicting a sequence of first-order transitions
assumes that at such transition the energy of a domain wall 
between the two neighboring phases is positive.
If, on the other hand, the domain wall energy turns out to be negative,
then the different phases want to mix and instead the true ground state 
has an inhomogeneous flux distribution, possibly modulated 
at some wavevector.  This interesting scenario, which is usually
believed to be less likely, has not been explored in the present work.
\end{thebibliography}
\end{document}